\documentclass{article}
\usepackage{spconf,amsmath,graphicx}

\usepackage{caption}
\usepackage{color}

\usepackage{subcaption}
\usepackage{graphicx}

\title{Enhancing Sound Texture In CNN-Based Acoustic Scene Classification}
\name{Yuzhong Wu, Tan Lee}
\address{Department of Electronic Engineering, The Chinese University of Hong Kong, Hong Kong SAR, China}
\begin{document}
\ninept 









\twocolumn

\maketitle
\begin{abstract}
Acoustic scene classification is the task of identifying the scene from which the audio signal is recorded. Convolutional neural network (CNN) models are widely adopted with proven successes in acoustic scene classification. However, there is little insight on how an audio scene is perceived in CNN, as what have been demonstrated in image recognition research. In the present study, the Class Activation Mapping (CAM) is utilized to analyze how the log-magnitude Mel-scale filter-bank (log-Mel) features of different acoustic scenes are learned in a CNN classifier. It is noted that distinct high-energy time-frequency components of audio signals generally do not correspond to strong activation on CAM, while the background sound texture are well learned in CNN. In order to make the sound texture more salient, we propose to apply the Difference of Gaussian (DoG) and Sobel operator to process the log-Mel features and enhance edge information of the time-frequency image. Experimental results on the DCASE 2017 ASC challenge show that using edge enhanced log-Mel images as input feature of CNN significantly improves the performance of audio scene classification.
\end{abstract}


\section{Introduction}
\label{sec:intro}

Large amount of multimedia information becomes easily accessible nowadays. The performance of speech and image recognition systems has been significantly improved with the use of deep neural networks and exploding amount of training data. Audio-related tasks, e.g., Acoustic Scene Classification (ASC) \cite{asc-1,Mesaros2016_EUSIPCO,asc-2}, Sound Event Detection (SED) \cite{sed-1,sed-2,sed-3} and Audio Tagging \cite{audio-tagging1,audio-tagging2,audio-tagging3,audio-tagging4}, have also received increasing attention in recent years. They have many real-world applications. For example, context-aware mobile devices could provide better responses to their users in accordance with the acoustic scene. A smart home-monitoring system could detect unusual incidences by using audio. An audio search engine is able to retrieve information efficiently from massive online recordings. 

Acoustic scene classification (ASC) is the process of identifying the type of acoustic environment (scene) where a given audio signal was recorded. It has been a major task in the IEEE AASP Challenge on Detection and Classification of Acoustic Scenes and Events (DCASE) since 2013. In the 2017 ASC challenge, most of the best-performing models were based on convolutional neural networks (CNN). Mun et al. \cite{asc2017-gan} addressed the problem of data insufficiency and proposed to use the Generative Adversarial Network (GAN) \cite{gan-invent} to augment training data. Han et al. \cite{Han2017} was focused on preprocessing of input features. Fusion of CNN models with preprocessed input features led to improved overall model performance. 

Despite the clearly demonstrated effectiveness of CNN-based models in the ASC task, there is little insight on how an audio scene is perceived in a CNN model. Whilst similar issue has been extensively explored in image classification. In \cite{visualize-cnn-deconv}, Zeiler \& Fergus used the De-convolutional Network \cite{deconv-net} to visualize and understand CNN. Springenberg et al. applied the guided backpropagation \cite{guided-backprop-all-conv-net} to obtain sharp visualization of descriptive image regions. The Class Activation Mapping (CAM) \cite{cam} was proposed as a means of highlighting the discriminative image regions for specific output classes in CNNs with global average pooling. Selvaraju et al. developed a generalized version of CAM, named the Gradient-weighted Class Activation Mapping (Grad-CAM) \cite{grad-cam}, which could be applied to a broader range of CNN models. 

The input of an audio classification model is usually a time-frequency representation extracted from the raw audio waveform. Among the various types of time-frequency representations, the logarithmic-magnitude Mel-scale filter bank (abbreviated as log-Mel) feature is widely adopted. Similar to spectrogram, a log-Mel feature is a visual representation of the frequency content of sounds as they vary with time. Given an audio signal with audible sound events such as ``bird singing'', ``speech'', ``applause'', these sound events can also be identified in the corresponding log-Mel feature based on their distinct visual patterns. From this perspective, we may call a log-Mel feature as an image. Visualization of CAM using the log-Mel ``image'' allows the comparison between machine perception and human interpretation. 



In this paper, we present an attempt to understand how CNN models learn to identify an acoustic scene from log-Mel feature representations. The investigation starts with benchmark systems with log-Mel features and different CNN models. The method of CAM is used to provide visualization of the CNN activation behavior with respect to input features. The observed CAMs for acoustic scene data suggest that CNN classification models tend to emphasize on the overall background sound texture of log-Mel input features, whilst individual sound events in the scene are of less importance. Hence we propose to use the Difference of Gaussian (DoG) and the Sobel operator to pre-process the log-Mel feature to make the background texture information more salient. We also use the method of background drift removing with medium filter as described in \cite{Han2017} as a comparison to our methods. These texture-enhanced features demonstrate an improved performance on ASC.


\section{BACKGROUND}
\label{sec:background}

\subsection{Class Activation Mapping}
\label{ssec:intro-cam}

The class activation mapping \cite{cam} highlights the class-specific discriminative regions in the input image. It can help understand the CNN behavior and visualize the internal representation of CNNs. It can also be used for weakly supervised object localization task. However, the CAM is only applicable to CNNs with global average pooling (GAP). Suppose we have a trained CNN network with global average pooling. There are $C$ output classes. The number of channels in the last convolutional layer is $K$. The point $(x,y)$ in the $k^{th}$ feature map before GAP as $f_k(x,y)$. The weight in the output layer is denoted as $w^c_k$, indicating the importance of the $k^{th}$ feature map for class $c$. Then the classification score of class $c$ (before the softmax) is given by
\begin{equation}
\label{eq:class-score}
y^c = \sum_{k} w^c_k \sum_{x,y}f_k(x,y) = \sum_{x,y}\sum_{k} w^c_k f_k(x,y).
\end{equation}
Based on equation \ref{eq:class-score}, the spatial elements of class activation map $M_c$ for class $c$ is given by
\begin{equation}
\label{eq:cam}
M_c(x,y) = \sum_{k} w^c_k f_k(x,y).
\end{equation}

The Gradient-weighted Class Activation Mapping (Grad-CAM) \cite{grad-cam} is a strict generalization of CAM. It replaces the weight of each activation map $w^c_k$ with the average gradient back-propagated to each feature map, which is given by
\begin{equation}
\label{eq:grad-cam}
\alpha^c_k = \frac{1}{Z} \sum_{i,j} \frac{\partial y^c}{\partial f_k(i,j)},
\end{equation}
where $Z$ is the number of pixels in the feature map. Notice that $f_k$ in Grad-CAM can be from any convolutional layers in CNN, not  limited to the last convolutional layer. Thus, the Grad-CAM can be applied to a larger variety of CNN models, such as those with fully connected layers (e.g. AlexNet, VGG).

In this paper, we propose to use Grad-CAM to analyze the trained CNN models for ASC task. Through empirical analysis of class activation maps w.r.t. the ground-truth scene classes, we argue that CNN models are more focusing on the overall background sound texture for classifying acoustic scenes. The distinct sound events (foreground) are usually of less importance in classification. 


\subsection{Sound Texture}
\label{ssec:intro-sound-texture}

Texture is described as an attribute that characterize spatial arrangement of pixel intensities in specific regions of an image. In the area of computer vision, texture analysis is a well studied topic \cite{6136524,Puig:2010:AFS:1823245.1823288,Chen:2006:MFS:1195775.1195785,Bhalerao03discriminantfeature}. 

For audio signal, the notion of ``sound texture'' has not been seriously discussed. An visual analogy of sound texture given by Saint-Arnaud et al. \cite{Saint-arnaud95analysisand} is that sound texture is like a wallpaper which has local structure and randomness, while from a large scale the fine structure characteristics must remain constant. There were a number of studies on sound texture modeling \cite{st-art-sound-texture-synthesis,sound-texture-modelling,sound-texture-gen}, and commonly mentioned sound textures refer to  wind, traffic, and crowd sounds.

In an acoustic scene, there exist various sound sources, which contribute to a mixture of diverse sound events. In audio recordings from acoustic scenes, persistent environment sounds with certain sound textures, e.g., crowd, traffic, form “background” of the scenes. Whilst certain sparsely occurred sound events, e.g., bird singing, human coughing, are more noticeable and could be regarded as distinct “foreground” sounds.

\subsection{Feature Preprocessing Methods}
\label{ssec:intro-methods-preprocess}

\subsubsection{Difference of Gaussian}
\label{sssec:intro-dog}

The Difference of Gaussian (DoG) is a well-known method of edge detection in image processing. Briefly speaking, the DoG filtering includes two steps: blurring an image using two Gaussian kernels of different standard deviations, and subtracting one blurred image from another to obtain the edge image. The purpose of Gaussian kernel is to suppress the high (spatial) frequency information (which serves as a low-pass filter). The value of standard deviation decides the range of frequency being suppressed. DoG essentially acts like a band-pass filter. It removes not only high (spatial) frequency noise, but also homogeneous regions in the image.

\subsubsection{Sobel Operator}
\label{sssec:intro-sobel}

The Sobel operator \cite{sobel} is commonly used for edge detection in computer vision. It comprises two $3\times 3$ convolution kernels, which 
are used to obtain the gradient approximations in the horizontal direction ($G_x$) and vertical direction ($G_y$). For an image $A$, we have 
\begin{equation}
\label{eq:sobelgxgy}
G_x =   \begin{bmatrix}
     +1 & 0 & -1 \\
     +2 & 0 & -2 \\
     +1 & 0 & -1 \\
   \end{bmatrix} * A,~~ G_y =   \begin{bmatrix}
     +1 & +2 & +1 \\
     0 & 0 & 0 \\
     -1 & -2 & -1 \\
   \end{bmatrix} * A.
\end{equation}
The gradient approximations in different directions can be combined as $G$, as the result of Sobel filtering:
\begin{equation}
\label{eq:sobelg}
G = \sqrt{G_x^2 + G_y^2} .
\end{equation}

\subsubsection{Removing Background Drift Using Medium Filter}
\label{sssec:intro-medfil}

Median filtering is useful in distinguish objects in an image with transitional background. By subtracting the medium-filtered image from the original one, the background drift is removed and those sharp changes (edges) are preserved \cite{medfil-bs}. For the ASC task, median filtering was found to be very effective in feature pre-processing \cite{Han2017}, though the determination of kernel size for optimal performance is not straightforward.

\section{ACOUSTIC SCENE CLASSIFICATION SYSTEM}
\label{sec:asc-system}

\subsection{System Design}
\label{ssec:sysdesign}
Experiments on scene visualization and classification are all based on the TUT Acoustic Scenes 2017 database \cite{Mesaros2016_EUSIPCO}. This database was adopted for the DCASE 2017 ASC challenge. It has two subsets: the development dataset (for model training and cross validation) and the evaluation dataset (for performance evaluation).

All audio samples in the dataset are $10$-second long. They are cut into $1$-second segments with $0.5$ second overlapping. Short-Time Fourier Transform (STFT) is applied to each of the $1$-second segments, with window length of $25$ms, window shift of $10$ms and FFT length of $2048$. $128$-dimension log-Mel filterbank features are derived from the FFT spectrum for each frame. Feature components of all frequency bins are normalized to have zero mean and unit variance based on training data statistics.

The CNN model receives the log-Mel feature image of a $1$-second segment as the input, and generates a classification score for the segment. The classification score for a $10$-second audio sample is obtained by averaging the segment-level scores.


\subsection{Model Structure}
\label{ssec:mdlstruct}

We examine the performance of two different CNN models. The CNN-FC model as detailed in Table \ref{table:cnnfc} is inspired by the AlexNet \cite{newalexnet} and VGG \cite{vgg} model. After the last convolutional layer, the feature maps are flattened to obtain the input for the fully connected layer.

The CNN-GAP model described in Table \ref{table:cnngap} is constructed by replacing the fully connected part in CNN-FC model with a global average pooling layer. Global average pooling (GAP) has been proven to be a good regularizer for CNNs in image classification \cite{netinnet}. GAP is also used in CNNs with audio input feature \cite{Sakashita2018,Dorfer2018,Zeinali2018,audioclassify-yzwu}. The same setup for training and testing is adopted for both models unless stated otherwise.


\begin{table}[]
\caption{The CNN-FC model structure. }
\label{table:cnnfc}
\centering
\begin{tabular}{lc}
\hline
   & Input 1x100x128                               \\\hline
1  & 3x3 Convolution (pad-1, stride-1)-64-BN-ReLu  \\
2  & 3x3 Max Pooling (stride-2)                    \\\hline
3  & 3x3 Convolution (pad-1, stride-1)-192-BN-ReLu \\
4  & 3x3 Max Pooling (stride-2)                    \\\hline
5  & 3x3 Convolution (pad-1, stride-1)-384-BN-ReLu \\
6  & 3x3 Convolution (pad-1, stride-1)-256-BN-ReLu \\
7  & 3x3 Convolution (pad-1, stride-1)-256-BN-ReLu \\
8  & 3x3 Max Pooling (stride-2)               		\\\hline
   & Flattening										\\
9  & Dropout (p=0.5)                       			\\
10 & Fully Connected (dim-2048)-BN-ReLU  			\\
11 & Dropout (p=0.5)                       			\\
12 & Fully Connected (dim-2048)-BN-ReLU				\\
13 & 15-way SoftMax                               \\\hline
\end{tabular}
\end{table}
 
\begin{table}[]
\caption{The CNN-GAP model structure. }
\label{table:cnngap}
\centering
\begin{tabular}{lc}
\hline
   & Input 1x100x128                               \\\hline
1  & 3x3 Convolution (pad-1, stride-1)-64-BN-ReLu  \\
2  & 3x3 Max Pooling (stride-2)                    \\\hline
3  & 3x3 Convolution (pad-1, stride-1)-192-BN-ReLu \\                 
4  & 3x3 Max Pooling (stride-2)                    \\\hline
5  & 3x3 Convolution (pad-1, stride-1)-384-BN-ReLu \\
6  & 3x3 Convolution (pad-1, stride-1)-256-BN-ReLu \\
7  & 3x3 Convolution (pad-1, stride-1)-256-BN-ReLu \\
8  & 3x3 Max Pooling (stride-2)                    \\\hline
9  & Global Average Pooling                        \\
10 & 15-way SoftMax                               \\\hline
\end{tabular}
\end{table}

\section{Visualization with Class Activation Maps}
\label{sec:visualizecam} 

For a given audio segment ($1$-second long in this study), the short-time log-MEL features could be viewed as a gray-scale image, with the x axis and the y axis representing time and frequency respectively. The image is combined with class activation maps for localizing the discriminative time-frequency regions. The activations are derived for the ground-truth scene class, and thus can be used to represent the input patterns learned by the CNN.

The proposed CAM visualization of audio segment is created by mixing 3 image components. The first component is the gray-scale log-MEL image. The time-frequency regions that positively influence the classification score of the ground-truth scene class are indicated by a semi-transparent image in red color.  The negative activations are viewed as another semi-transparent image of blue color. Being different from \cite{grad-cam}, both positive and negative activations are included for the observation of acoustic scene features.

\begin{figure}[h]
\centering
\begin{subfigure}[b]{0.5\textwidth} 
   \includegraphics[width=1\linewidth]{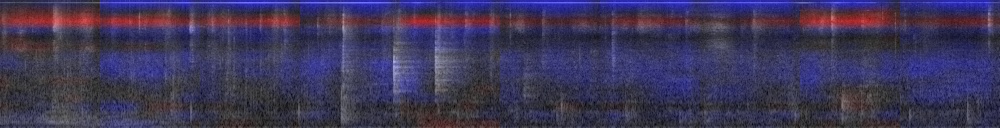}
   \caption{metro station}
   \label{fig:cam-metro} 
\end{subfigure}
\begin{subfigure}[b]{0.5\textwidth}
   \includegraphics[width=1\linewidth]{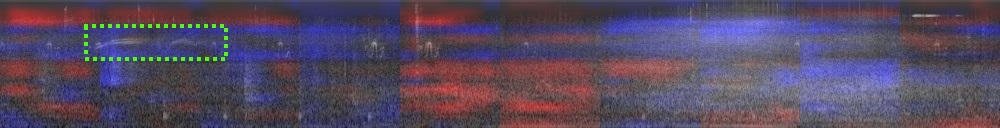}
   \caption{residential area}
   \label{fig:cam-resident}
\end{subfigure}
\begin{subfigure}[b]{0.5\textwidth}
   \includegraphics[width=1\linewidth]{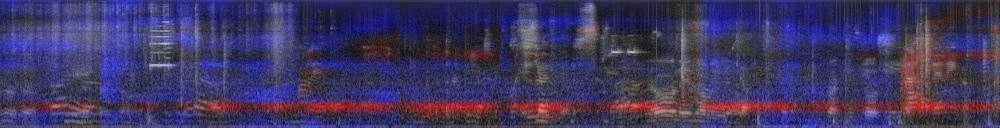}
   \caption{train}
   \label{fig:cam-train}
\end{subfigure}
\caption{CAM visualizations w.r.t the ground-truth scene classes. They are derived from the CNN-FC model with log-Mel input. The $3$ audio samples are recorded in (a) metro station, (b) residential area  and (c) train respectively. High energy regions (distinct sound events) are not strongly activated. It seems that the model is trying to learn the texture of background sounds.}
\label{fig:cams-cnnfc}
\end{figure}

Figure \ref{fig:cams-cnnfc} gives a few examples of gradient-weighted CAM visualization derived from the $7^{th}$ layer feature maps in CNN-FC model. These $10$-second audio samples are from the training set of DCASE 2017 dataset. In Figure \ref{fig:cam-resident}, the white horizontal lines during the first $3$ seconds (inside the green dashed line rectangle) are ``bird singing'' sounds. It is noted that these distinct sound events are not associated with strong positive (red) or negative (blue) activation. In other words, in CNN classification, these sounds are not regarded as representative patterns for the residential area scene.

\begin{figure}[h]
\centering
\begin{subfigure}[b]{0.5\textwidth} 
   \includegraphics[width=1\linewidth]{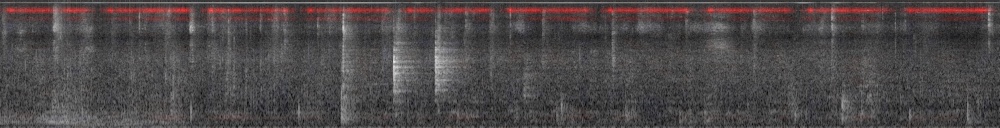}
   \caption{metro station}
   \label{fig:cam-cnngap-metro} 
\end{subfigure}
\begin{subfigure}[b]{0.5\textwidth}
   \includegraphics[width=1\linewidth]{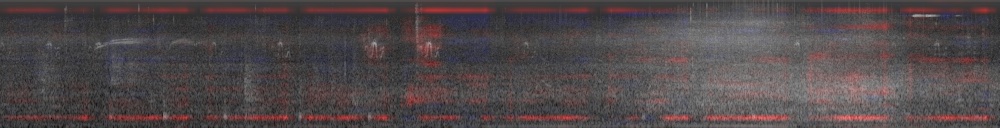}
   \caption{residential area}
   \label{fig:cam-cnngap-resident}
\end{subfigure}
\begin{subfigure}[b]{0.5\textwidth}
   \includegraphics[width=1\linewidth]{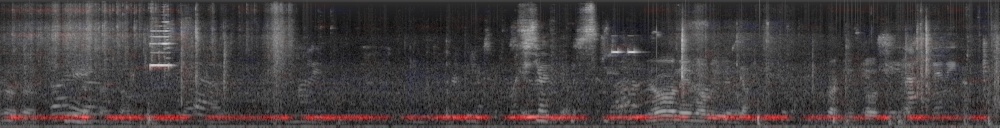}
   \caption{train}
   \label{fig:cam-cnngap-train}
\end{subfigure}
\caption{CAM visualizations w.r.t the ground-truth scene classes. They are derived from the CNN-GAP model with log-Mel input. The audio samples are the same as those in Figure \ref{fig:cams-cnnfc}. Different from CNN-FC model, there is little negative activation for CAMs derived from CNN-GAP model. }
\label{fig:cams-cnngap}
\end{figure}

Figure \ref{fig:cams-cnngap} shows the examples of CAM visualization derived from the $7^{th}$ layer feature maps in CNN-GAP model. As we can see from these examples, the magnitudes of positive activations are much higher than those of negative activations. The activations are concentrated on a few frequency bins, unlike the CNN-FC model. In addition, the eye-catching bright lines (foreground sounds) in the log-MEL images are associated with low activation intensity. This suggests that the CNN-GAP model performs classification based on the background ``texture'' of input image.


It frequently happens that the regions of distinct sound events in the log-MEL images have small activation intensity, which might be counter-intuitive. However, further investigation is needed to find out if these sound events are really trivial for classification, or it is because the CNN models fail to learn these patterns.





\section{Enhancing The Edge Information in log-Mel Images}
\label{sec:enhancing-edge}

\subsection{Edge-Enhanced Features}
\label{ssec:edge-enhanced-features}

\begin{figure}[h]
  \centering
  \includegraphics[width=0.8\linewidth,keepaspectratio]{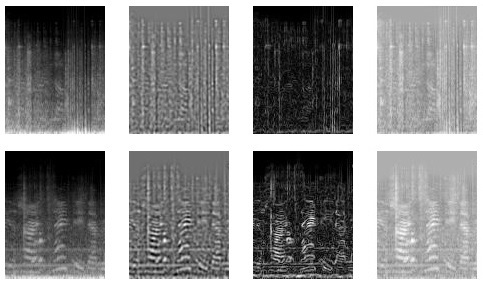}
  \caption{Illustration of edge-enhanced input features for CNNs. (From left to right) First column: Original log-Mel image; second column: edge-enhanced images with DoG; third column: edge-enhanced images with Sobel operator; fourth column: background drift removed images with medium filter. }
  \label{fig:features}
\end{figure}

We propose to use DoG and Sobel operator to enhance the edge information in the input images, making the background texture more salient. Figure \ref{fig:features} gives a few examples of edge-enhanced images and the corresponding unenhanced ones.

To obtain the DoG enhanced image, we apply Gaussian filter with standard deviation $1$ to the original log-Mel image. Then we apply another Gaussian filter with standard deviation $\sqrt{2}$ on the original image to obtain another blurred image. Subtraction between these two blurred image gives the result of DoG. DoG is able to remove high spatial-frequency components and homogeneous regions of images.

The result of Sobel operator is an image with pixel values being equal to the gradient magnitudes of the respective pixels in the original image. Comparing to DoG, the images enhanced by Sobel operator have more fine-grained texture.

The above mentioned edge-enhanced images are compared to the one obtained by median filtering. The kernel size of medium filter is empirically set to $(51,7)$ where $51$ refers to time frames (about $0.5$ second) and $7$ refrs to frequency bins. It is noted that median filtering process has a high computation cost, i.e., requiring to calculate the median of each $(51,7)$ input window for each output pixel.

\subsection{Evaluating the Edge-Enhanced Features}
\label{ssec:evaluate-edge-enhancement}

Table \ref{table:experiment-results} shows the accuracy (averaged over 3 trials) for different types of input features. The ``LogMel-128'' means the 128 dimensional log-Mel feature, which is considered as benchmark feature. ``DoG'' and ``Sobel'' refer to the DoG enhanced and Sobel operator enhanced LogMel-128 features, respectively. ``Medium'' refers to the background-drift-removed LogMel-128 feature (using medium filter). The baseline system accuracy is provided by the DCASE 2017 ASC challenge \cite{dcase2017-baseline}. It can be seen that applying edge-enhancement techniques leads to significant improvement of classification performance. We also check the CAM visualizations of CNN models with the edge-enhanced input images, and the observations in Section \ref{sec:visualizecam} are still valid. While the CNN-FC model and CNN-GAP model are different in visualized patterns of CAM, they show similar performance given the same input feature.

While the performance of using ``DoG'' feature is not as good as the ``Medium'' feature, DoG is computationally much more efficient than median filtering. For edge-enhanced features from 100 log-Mel images of size $(1000,128)$, computing ``Medium'' features takes $272.02$ seconds with kernel size $(51,7)$ in our computer. If the kernel size is changed to $(3,3)$, the computation time is $5.7$ seconds. On the other hand, applying DoG and Sobel operator takes $0.46$ and $0.30$ second respectively.

\begin{table}[]
\centering
\caption{Experiment results for CNN-FC and CNN-GAP models trained with different input features. The evaluation data in TUT Acoustic Scenes 2017 database is used for evaluation. Table element is the overall classification accuracy (averaged over 3 trials). The accuracy for the baseline setup is from the DCASE 2017 ASC challenge \cite{dcase2017-baseline}. }
\label{table:experiment-results}
\begin{tabular}{|l|l|l|l|}
\hline
Feature\textbackslash Model          & CNN-FC & CNN-GAP & Baseline\\ \hline
Baseline   & - 		  & -               & $0.610$ \\ \hline
LogMel-128 & $0.658$    & $0.681$     & -       \\ \hline
DoG        & $0.720$    & $0.722$     & -       \\ \hline
Sobel      & $0.701$    & $0.716$     & -       \\ \hline
Medium  & $0.757$    & $0.754$     & -       \\ \hline
\end{tabular}
\end{table}



\section{CONCLUSION}
\label{sec:conclusion}
In this paper, we illustrate the use of class activation mapping for analysis of CNN behavior towards audio features. We find that the distinct sound events in log-Mel features are usually not regarded as representative patterns of acoustic scenes. Regarding ASC task as a sound texture classification problem, we use the DoG, Sobel operator and background drift removing to enhance the edge information in the log-Mel image. Using these methods, the model performance is improved significantly compared to the benchmark.



 
\bibliographystyle{IEEEbib}
\bibliography{refs}

\end{document}